# Power Scalability as a Precise Concept for the Evaluation of Laser Architectures

Rüdiger Paschotta

*RP Photonics Consulting GmbH, Kurfirstenstr. 63, 8002 Zürich, Switzerland*
*Paschotta@rp-photonics.com*

**Abstract:** This paper establishes power scaling of lasers as a clearly defined concept, based on a power scaling procedure which must satisfy various criteria. It is demonstrated that this concept creates useful insight particularly for the evaluation of the future performance potential of different laser architectures, and for identifying technological aspects which will need to be modified for generating higher powers. It turns out that some aspects (such as e.g. thermal lensing in thin disk lasers) can have rather benign scaling properties, not causing problems even at very high power levels, while other aspects can become essential even if they initially may have appeared to be insignificant.

PACS: 42.55.Rz, 42.55.-f, 42.55.Ah

In recent years, enormous performance enhancements have been achieved with laser systems of various kinds. One aspect of particular interest is the output power, often with the additional requirement of a good beam quality, as needed for strong beam focusing over a reasonably long working distance. Remarkable advances have been reported for active fiber devices, namely for fiber lasers and amplifiers operating in the 1-µm wavelength regime. While a few tens of watts were considered a high output power of such systems around 1990, well over 1 kW is nowadays possible with nearly diffraction-limited beam quality [1]. Other laser architectures, based e.g. on thin disk lasers [2] or slab lasers [3], have also exhibited strong performance enhancements. In many cases, such advances have been labeled as the result of *power scaling*.

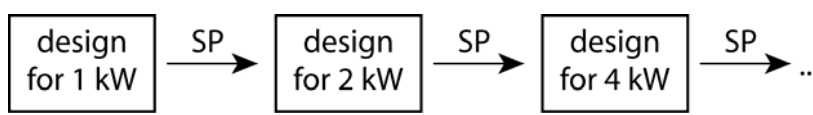

Fig. 1: Repeated application of a scaling procedure (SP) to a power-scalable design allows to realize devices in a wide range of output powers, without hitting certain limiting effects

In many areas of modern technology, including e.g. computing, data storage systems, telecommunications, and manufacturing, the concept of *scalability* plays an important role particularly in the context of estimating and realizing growth potentials. For example, a type of telecommunications network or a data storage system is called *scalable* when its data throughput or data capacity can be expanded in a wide range simply by adding hardware of the same type, rather than e.g. by developing improved hardware components or by utilizing additional inventions. In more quantitative terms, the essential idea is that certain technological approaches or system architectures, in contrast to others, allow the formulation of so-called *scaling procedures*, i.e., systematic methods to transform an existing and working device or system into another one with e.g. 2 times or 10 times the performance in some respect (see Fig. 1). Ideally, such a scaling procedure could be applied indefinitely, thus allowing to reach arbitrarily high performance levels. While such ideal conditions can hardly be found anywhere in the real world, it is sensible to demand from a scaling procedure that it allows the realization of performance figures at least within some large parameter range. Furthermore, this expansion of performance should be possible without encountering major problems, i.e., without relying on additional inventive steps for the mitigation of certain limiting effects. If a certain technological route can be shown to exhibit such kind of scalability, one may hope that further scaling can be extrapolated into not yet explored regions of even higher performance levels.

In the context of laser technology, the terms "power scaling" and "scalability" are nowadays very widely used – but usually without reflecting at all the meaning of true scaling. Indeed it has become commonplace to use such terms even in situations where a repeatable method for performance improvements is definitely not involved. As an example, the pump power and thus possibly the output power of a solid state laser can be doubled by polarization coupling two laser diodes. This, however, results in an unpolarized beam, so that the method can obviously be applied just once. In other cases, laser systems are operated with borderline intensity levels, and the methods used for somewhat pushing away the experienced limiting effects are often far from a systematic repeatable procedure as required for scalability.

Although the improper use of such terms is widely and irreversibly spread in the literature, the concept of scalability can be revitalized in the context of laser technology. In that way, the essential and helpful thoughts behind the concept of scaling could be better utilized for this branch of technology. Exactly that is the goal of this article.

# 1    Criteria for True Power Scalability

What is needed is (a) a reasonable terminology and (b) a concrete application of the concept of scaling to that particular technology. Task (a) is easily solved by using the extended terms "true power scaling" and "true power scalability" when referring to methods or devices which satisfy a meaningful set of requirements. Task (b) is then to work out these requirements in such a manner that the application of the scaling concept in laser technology becomes most fruitful, i.e., practical, clear, and not biased toward any particular technology. In the following sections, this paper shows how the concrete application to various laser architectures can start, without however going into great detail with any particular type of laser.

True power scaling of lasers should obviously be based on a well-defined *scaling procedure*, which describes how to transform an existing and working laser design into another design with e.g. twice the output power. The most basic requirements which are reasonable to demand are:

(1) The procedure is *systematic and well defined*, including quantitative aspects. For example, the form of such a procedure could be: *For doubling the output power, double the pump power, apply it to twice the mode area in the laser gain medium, reduce parameter X by one half while keeping parameter Y constant.*

(2) The procedure must be *repeatable*. It would not be sensible to demand *indefinite* repeatability, which can not be achieved with any real physical system. However, it should be possible to vary the output power in a large range.

While requirement (1) is quite obvious and clear, requirement (2) needs further refinement. Thinking about frequently encountered issues in the given technological context, it seems reasonable to more closely specify requirement (2) as follows:

(2a) Application of the scaling procedure *should not spoil essential features* of the laser, such as e.g. its power conversion efficiency or beam quality. What is essential may however depend on the concrete circumstances. For example, the constraint of preserving the beam quality is often important, but may be dropped in some cases where beam quality is not relevant for the targeted application.

(2b) The procedure must be devised so that enhanced laser designs do not rely on increasingly critical specifications for the components to be used in the laser. For example, it must not demand the use of pump laser diodes with arbitrarily increased brightness.

(2c) The procedure must not make one or more central challenges, such as e.g. the handling of high optical intensities or strong thermal effects, significantly more severe as the power is increased.

In the following, I will apply the found criteria to some concrete laser architectures of high importance. The intention is *not* to provide a comprehensive discussion and comparison of all important types of high power lasers, but rather to demonstrate how a precise use of the concept helps to gain a deeper insight into power scaling issues. This can then form the basis for fruitful further work by other authors on various laser architectures.

## 2 Discussion of Various Laser Architectures

### 2.1 Diode-Pumped Solid-State Bulk Lasers

The still most common type of solid state laser is the bulk laser, where the gain medium is a bulk crystal (or a piece of glass), typically a rod which is doped with rare earth ions (e.g. $Nd^{3+}$ or $Yb^{3+}$), pumped with laser diodes, and cooled from the side. Substantial power enhancements have been achieved within the last twenty years. These are to a large extent based on the availability of laser diodes with higher and higher power and brightness, and on the refinement of many technical details. Current high power rod lasers – at least those with high beam quality – are actually no more limited by available pump power, but by thermal effects in the gain medium, such as thermal lensing (associated with strong aberrations) and stress fracture. Increasing the pump and laser beam radius in the laser crystal does reduce the focusing power of the thermal lens, but a larger laser mode is also more sensitive to lensing effects, so that overall the impact of lensing is not reduced. The use of a longer crystal reduces the risk of stress fracture, but does not reduce the total amount of lensing and aberrations. For such reasons, thermal effects in laser rods inevitably become more challenging at higher powers. Currently achieved power levels are near that limit, although various technical measures (using e.g. smoother pump intensity distributions, doping gradients, intracavity Faraday rotators, etc.) help to stretch the accessible power range somewhat further. This means that there appears to be no scaling procedure taking into account these effects, and power scalability thus exists only up to the thermal limit which is reached already.

The technology of slab lasers allows for higher powers. However, this concept quite naturally enforces the use of slabs with large aspect ratios, and it is challenging to efficiently extract the optical power with high beam quality, e.g. using strongly elliptical laser modes or folded beam geometries. In fact it appears that power scaling attempts with slab lasers [3] are not targeting diffraction-limited beam quality, and thus take place in a somewhat different discipline. It seems, however, that a relatively well defined scaling procedure can be formulated within that discipline. This should at least allow for very high output powers without hitting the limits of excessive temperatures or mechanical stress in the laser crystal [3], and that progress does not depend on arbitrarily improved laser diodes for pumping. For some range of applications, the achievable beam quality may be quite sufficient.

One special kind of bulk laser, the thin disk laser [2], is different from other bulk lasers in terms of scalability. Due to its geometry, with heat removal from the active disk essentially in the beam direction rather than in a transverse direction, the dioptric power of the thermal lens is not only reduced, but (more importantly) also acquires a different scaling behavior, comparing with that e.g. of side-cooled rod lasers. At least with a simplified model, assuming a purely one-dimensional heat flow in the disk into a heat sink with constant surface temperature, and regarding only thermal effects via the temperature dependence of the refractive index (but not stress effects), it can easily be shown that the dioptric power of the thermal lens scales in inverse proportion to the mode area, if the mode area and pump area are increased in proportion to the pump and output power. That inverse dependence then cancels the effect of the smaller stability range (with respect to dioptric power) of a resonator with a larger mode size on the disk. In effect, the width of the stability zone of the resonator with respect to pump power scales in proportion to the maximum pump power. (The mentioned cancellation would not occur e.g. in a rod laser with cooling in radial direction, where the dioptric power stays constant when the mode area is increased in proportion to the pump power. Here, the width of the stability range thus becomes a smaller and smaller percentage of the pump power for higher power devices.) Similar results can be derived for thermally induced aberrations in the disk; the phase changes related to those do not become stronger for devices operating at higher power levels. Overall the challenge of thermal lensing effects thus does *not* become more severe at higher power levels. This is actually no more true once stress-induced effects become important. While stress in principle sets a limit to the power scalability, it can be quite effectively mitigated with various means, e.g. by using a cap of undoped YAG on top of the doped part of the disk. There is also a number of technical difficulties related to exactly how the disk is fixed on its cooling finger, ensuring good thermal contact while avoiding excessive mechanical stress with and without pumping. Such issues, however, can be addressed with certain soldering schemes, and will probably not prevent the scaling of thin disk lasers well beyond 10 kW with diffraction-limited beam quality. The highest demonstrated powers have so far actually been associated with somewhat reduced beam quality, because the above mentioned issues have not yet been systematically worked out, and probably also because there was no pressing need to further improve the already good beam quality for the intended applications e.g. in remote welding. At this stage it appears that this laser architecture is essentially power scalable up to the multi-kW regime, even with very high beam quality, even though some not entirely trivial problems are involved, e.g. exactly how to mitigate the stress effects for diffraction-limited beam quality at multiple kilowatts [4]. Essentially, the scaling procedure is as follows: for doubling the output power, double the pump power and the mode area on the disk, while keeping the disk thickness and the output coupler transmission constant. This procedure preserves the optical intensity as well as the laser gain and (approximately) the disk temperature, and it scales the width of the resonator's stability region in terms of pump power in proportion to the output power, as far as stress effects can be disregarded. By working with a small disk thickness (and correspondingly high doping density), stress effects are kept weak in a wide range of powers. The required pump brightness is not increased by the scaling, nor the required number of pump passes through the disk (assuming a constant disk thickness). Perhaps surprisingly, such scaling aspects (e.g. related to dioptric powers) appear not to have been worked out in detail in the literature, despite the great interest in this technology.

Note that the power scaling of *any* solid state laser should involve increases of the mode area (at least in the laser crystal) roughly in proportion to the output power, since the optical intensities have to be limited. (In fact, the intracavity laser intensity in the crystal is almost always just several times higher than the saturation intensity.) For bulk lasers (with unguided light propagation), this has important consequences. If the mode area on all optical components has to be scaled up while keeping the resonator length limited, the effect of diffraction becomes weaker and weaker. This introduces various problems, in particular that of resonator stability, because changes of focusing conditions will result in large changes of mode area, and the alignment sensitivity becomes higher. (Telescopic resonators only partly meet this challenge, since they lead to higher intensities in parts of the resonator.) Also, the Gouy phase shifts then tend to be small, and this leads to excessive coupling from the fundamental mode to higher-order transverse modes, in effect compromising the beam quality [5]. There are thus some challenges to be expected for diffraction-limited emission in the multi-kW regime: laser resonators appear not to be strictly power scalable, at least for a limited resonator length.

To a limited extent, power scaling is possible by increasing the number of laser heads within the resonator as well as the output coupler transmission. The output power is then increased without increasing the optical intensities in the laser heads. The impacts of thermal effects in the laser heads do not necessarily add up; a possible solution is a periodic resonator [6]. However, this scaling method appears not to work over a very large range of powers, and is normally used only with a quite moderate number of laser heads.

## 2.2 Fiber Lasers and Amplifiers

The use of waveguide structures, e.g. of optical fibers, constitutes another method of coping with strong thermal effects in the laser gain medium. As long as the impact of the guiding structure dominates over that of thermal effects – which can be the case even at quite high power levels – thermal lensing is not detrimental.

Another issue is to launch sufficient pump power into a fiber device. While launching only into the two fiber ends sets a limit to the achievable power, there are enhanced techniques for coupling pump power into double-clad fibers also at intermediate locations. Although previous advances of fiber lasers have often partly relied on the increased brightness of pump diodes, advanced schemes for injecting pump light do not introduce scaling limitations.

However, increasing power levels have lead to substantially increased optical intensities in the fiber core, which can not only cause damage but also introduce detrimental nonlinear effects such as stimulated Raman or Brillouin scattering. Although e.g. Raman scattering can be suppressed by using a fiber design with high losses for the Raman-shifted wavelength region, and Raman as well as Brillouin scattering can sometimes be reduced by limiting the fiber length, at least damage phenomena set a definite limit to the optical intensity. Therefore, increased powers ultimately require an increase of mode area. Fiber mode areas have already been increased by more than an order of magnitude, and it appears that there is not much room for further increases as long as single-mode guidance for diffraction-limited beam quality is demanded. The problem is essentially that the optical guidance is obtained from a balance of waveguide effects with diffraction, and this balance becomes more and more sensitive to other influences for large mode areas, where diffraction is very weak. Some schemes allow for somewhat larger mode areas with multimode fibers while still essentially preserving the high beam quality, but this potential also appears to be fairly limited, particularly by mode coupling.

As the optical intensities in high power fiber devices are already close to the limits set by nonlinear effects or even fiber damage, it becomes apparent that the output power per fiber with diffraction-limited beam quality can not be increased far beyond the currently achieved level of a few kilowatts, at least with the known type of fiber laser and amplifier technology. In other words, simple extrapolation of recent advances would produce wrong results.

Power scalability of fiber devices may be claimed up to the level of a few kilowatts. However, the most recent advances have largely not been the result of true power scaling, but rather of a combination of several non-repeatable factors, in particular increased power and brightness of the pump diodes, optimized cladding-pumped fibers, and designs with a more optimized balance of various limiting effects. A scaling procedure appears not to exist beyond the already realized mode areas (which would definitely have to be increased further), unless the beam quality can be sacrificed.

## 2.3 Optically Pumped VECSELs

Most semiconductor lasers are based on waveguide structures which face similar types of scaling limitations as fiber devices: the limited achievable mode areas. A relatively recent concept, namely the optically pumped vertical external-cavity surface-emitting laser (VECSEL) [7], removes this problem and can achieve true power scalability when combined with proper cooling techniques, such as using a quite thin semiconductor gain structure [7] and/or a heat spreader attached to the top surface [8]. (Without such measures, the longitudinal heat flow is not sufficient for cooling.) The core of the scaling procedure is then to increase the pumped area in proportion to the pump and output power. This limits the optical intensities as well as the temperature excursion – just as in a thin disk laser, exploiting an essentially longitudinal heat flow. Indeed, such devices have been demonstrated with output powers between a few hundred milliwatts and tens of watts [9], all with close to diffraction-limited beam quality. True power scalability has been the basis for such huge increases within just a few years.

Note that optical pumping is an essential prerequisite for scalability, since no scaling procedure for realizing arbitrarily large electrically pumped gain areas is known to date. While the achievement of output powers of the order of 1 W [10] from electrically pumped devices is certainly impressive, the lack of scalability contradicts the possible impression that electrically pumped VECSELs will reach the same high output powers as optically pumped devices, only perhaps a few years later.

## 2.4 Beam Combining

A scaling procedure may avoid the limitations of a single laser crystal or fiber by beam combining, i.e., by scaling the number of such components which contribute to a combined optical output [11]. There are several techniques of scalable beam combining with the potential to preserve the beam quality: wavelength division multiplexing (e.g. with dichroic mirrors), coherent combining on partially transparent mirrors, and transverse coherent combining. In the latter case, $N$ diffraction-limited modes are coherently combined to form a mode with $\approx N$ times the area and $\approx N^{-1/2}$ times the divergence angle.

The limits for such scaling procedures are often "soft limits" arising from practical issues, such as stability of mechanical alignment and relative optical phase, or simply from the large number of components and the complexity of stabilization techniques. Even though a solution with a single laser head or fiber has practical advantages in such respects, beam combining methods might turn out to provide the only viable path for extremely high output powers. It seems that hard limits for this type of power scaling are relatively far away.

## 3  Conclusions and Discussion

Substantial insight can be gained by applying the concept of scaling to the area of high power lasers. I have shown that power scalability in a meaningful sense must be based on a scaling procedure, which should satisfy several requirements, all essentially arising from basic demands of clarity and repeatability. Where true power scalability exists, the laser performance can be increased within a wide range of output powers without relying on arbitrarily improved components (e.g. pump diodes with higher brightness), new materials (e.g. with higher damage threshold) or additional inventive steps (e.g. to keep thermal effects under control).

Of course, every technology will finally encounter some limits, often set by certain effects which are not important in a wide range of parameters but become serious once a certain parameter region is reached. Such limiting effects are e.g. thermal lensing (including aberrations) in many bulk lasers (although to a much weaker extent in thin disk lasers), while optical nonlinearities and material damage are essential in fiber devices. For judging the potential of different laser architectures for further performance improvements, it is vital to precisely understand (a) the physical details of limiting effects, (b) their impact on the further development ("how they scale"), and (c) available options to mitigate or eliminate those effects with a strong impact without introducing or exacerbating other problems. Particularly for issue (b), it is essential to recognize the existence (or non-existence) of a scaling procedure which takes into account the potentially limiting effects. Fig. 2 (a) illustrates how an essentially damaging effect may eventually become too strong in a not power-scalable scheme, while Fig. 2 (b) shows such a dependence for a power-scalable architecture, where a potentially detrimental effect may be present from the beginning but does not become more severe as the power is increased. Examples are e.g. thermal lensing in a solid state rod laser, becoming more and more serious as in Fig. 2 (a), and thermal lensing in a thin disk laser, which increases in a much more benign way as in the upper curve of Fig. 2 (b). There may, however, be other effects which are initially weak but can eventually become limiting for higher powers, shown as the lower curve in Fig. 2 (b). An example for this are stress effects in a thin disk laser.

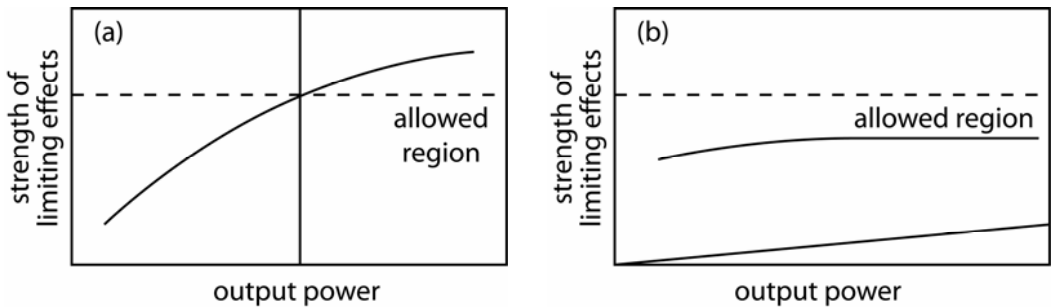

Fig. 2: Dependence of the strength of some possibly detrimental effects on the output power for two different laser designs. Power scalability as in case (b) implies that all effects stay under control in a wide range of powers, but scalability may end at even higher powers where new effects can become limiting (as indicated by the lower curve). For example, thermal lensing can behave as in diagram (a) for a rod laser, but as in the upper curve of (b) for a thin disk laser

As another example, it is interesting to review the development of *passively mode-locked* high power thin disk lasers. As semiconductor saturable absorber mirrors (SESAMs), which are used for mode locking, also have the geometry of a thin disk, initial concerns about absorber damage were unfounded, and a first demonstration already lead to an average output power of 16 W in 0.7-ps pulses [12]. The further development for up to 80 W average power with essentially the same pulse duration [13] did not make any SESAM issues more severe, and the optical intensity on the SESAM was not higher than in usual low-power devices. However, that development met a new challenge which had been insignificant at lower power levels: that of dispersion compensation. Chirped mirrors as used for that purpose exhibit higher losses than normal highly reflecting laser mirrors, so that somewhat surprisingly thermal effects on such mirrors eventually become more important than thermal effects in the saturable absorber. This is another case where the scalability is limited by an effect which initially appeared to be insignificant, and scaling considerations elucidate the situation. Of course, it is quite possible to develop chirped mirrors e.g. on thinner substrates (cooled from the back side) in order to restore scalability at this point, but it is remarkable that the development of novel SESAM types is *not* required due to their different scaling properties.

These examples demonstrate that it is essential to understand how certain influences scale with the required output power, and not just how strong they are for a particular design. Quite obviously, such considerations go far beyond a simple extrapolation of previous technological advances, and thus can substantially reduce the uncertainties concerning future developments. It should also be emphasized that the investigation of scaling aspects can not only be applied to whole laser architectures, but also to isolated aspects, such as e.g. dispersion compensation in high power mode-locked lasers as discussed above. Such considerations allow to evaluate the viability of certain techniques or components for higher power levels before these parameter regions are experimentally explored.

A question of particular current interest is that about laser technologies for multi-kilowatt diffraction-limited output. No technology known so far appears to allow for true scaling e.g. beyond the 10-kW level with diffraction-limited beam quality. Some laser architectures, e.g. of rod lasers, are already limited by thermal effects at much lower power levels, while fiber lasers and amplifiers are hitting critical intensity levels at a few kilowatts when diffraction-limited beam quality is demanded. Thin disk lasers and optically pumped VECSELs appear to be closest to truly power-scalable architectures, with a well-defined scaling procedure and no apparent very hard limits, despite some nontrivial technical issues. With somewhat relaxed beam quality requirements, slab lasers also offer a strong potential. The greatest potential, however, appears to be offered by the concept of beam combining, which will certainly allow for tens or hundreds of kilowatts.

Finally, it is important to note that the feature of power scalability can strongly depend on the given conditions. For example, power scalability of some laser architectures (e.g. slab lasers and fiber lasers) can immediately be strongly extended by dropping the requirement of preserving the beam quality. Also, a pulsed system (e.g. with nanosecond pulses) will much sooner run into nonlinear effects than a continuous-wave device, and may thus exhibit different limiting factors. While such aspects definitely make the classification of laser systems more complex, they do not question the general usefulness of the concept of power scalability, in particular for judging the potentials for further improvements.

**References**

[1] Y. Jeong, J. K. Sahu, D. N. Payne, J. Nilsson, "Ytterbium-doped large-core fiber laser with 1.36 kW continuous-wave output power", Opt. Express 12 (25), 6088 (2004)

[2] A. Giesen, H. Hügel, A. Voss, K. Wittig, U. Brauch, H. Opower, "Scalable concept for diode-pumped high-power solid state lasers", Appl. Phys. B 58, 363 (1994)

[3] J. M. Eggleston, T. J. Kane, K. Kuhn, J. Unternahrer, R. L. Byer, "Periodic resonators for average-power scaling of stable-resonator solid-state. lasers", IEEE J. Quantum Electron. QE-20, 289 (1984), and T. J. Kane, J. M. Eggleston, R. L. Byer, "The slab geometry laser – Part II. Thermal effects in a finite slab", IEEE J. Quantum Electron. QE-21, 1195 (1985)

[4] A. Giesen, private communication

[5] R. Paschotta, "Beam quality deterioration of lasers caused by intracavity beam distortions", Opt. Express 14 (13), 6069-6074 (2006)

[6] J. M. Eggleston, "Periodic resonators for average-power scaling of stable-resonator solid-state lasers", IEEE J. Quantum Electron. 24 (9), 1821 (1988)

[7] M. Kuznetsov, F. Hakimi, R. Spaque, A. Mooradian, "High-power (>0.5-W CW) diode-pumped vertical-external-cavity surface-emitting semiconductor lasers with circular $TEM_{00}$ beams", IEEE Photonics Technol. Lett. 9 (8), 1063 (1997)

[8] K. S. Kim, J. R. Yoo, S. H. Cho, S. M. Lee, S. J. Lim, J. Y. Kim, J. H. Lee, T. Kim, Y. J. Park , "1060-nm vertical-external-cavity surface-emitting lasers with an optical-to-optical efficiency of 44% at room temperature", Appl. Phys. Lett. 88, 091107 (2006)

[9] J. Chilla, S. D. Butterworth, A. Zeitschel, J. P. Charles, A. L. Caprara, Mr. K. Reed, L. Spinelli, "High power optically pumped semiconductor lasers", Proc. SPIE 5332, 143 (2004)

[10] J. G. McInerney, A. Mooradian, A. Lewis, A. V. Shchegrov, E. M. Strzelecka, D. Lee, J. P. Watson, M. K. Liebman, G. P. Carey, A. Umbrasas, C. A. Amsden, B. D. Cantos, W. R. Hitchens, D. L. Heald, and V. Doan, "Novel 980-nm and 490-nm light sources using vertical-cavity lasers with extended coupled cavities", Proc. SPIE 4994, 21 (2003)

[11] T. Y. Fan, "Laser beam combining for high-power, high-radiance sources", IEEE J. Sel. Topics in Quantum Electron. 11 (3), 567 (2005)

[12] J. Aus der Au, G. J. Spühler, T. Südmeyer, R. Paschotta, R. Hövel, M. Moser, S. Erhard, M. Karszewski, A. Giesen, U. Keller, "16.2 W average power from a diode-pumped femtosecond Yb:YAG thin disk laser", Opt. Lett. **25** (11), 859 (2000)

[13] F. Brunner, E. Innerhofer, S. V. Marchese, T. Südmeyer, R. Paschotta, T. Usami, H. Ito, U. Keller, "Powerful RGB laser source pumped with a mode-locked thin disk laser", Opt. Lett. **29** (16), 1921 (2004)